\documentstyle[epsf]{aa}
\begin{document}

\newcommand{\gsim}{\hbox{\rlap{$^>$}$_\sim$}}
%
   \title{The Early Afterglow}

   \author{Re'em Sari  \inst{1} \and Tsvi Piran \inst{2}
          }
   \institute{
              Theoretical Astrophysics, California Institute of Technology, 
              Pasadena, CA 91125, USA. \\
       \and   Racah Institute of Physics, The Hebrew University, 
              Jerusalem 91904, Israel , and \\
              Department of Physics, Columbia University, 
              New York, NY 10027, USA}

   \date{Received December 15, 1996; accepted }

   \maketitle

   \begin{abstract} We calculate the expected spectrum and light
   curves of the early afterglow.  For short GRBs the peak of the
   afterglow will be delayed, typically, by a few dozens of seconds
   after the burst.  The x-ray and $\gamma$-ray characteristics of
   this delayed emission provide a way to discriminate between late
   internal shocks emission (part of the GRB) and the early afterglow
   signal.  Detection of this delayed emission will prove the internal
   shock scenario as producing the GRB, and will pinpoint the initial
   Lorentz factor $\gamma_0$.  In the optical band, the dominant
   emission arises from the reverse shock. This shock, carries an energy
   comparable to that of the forward shock. It radiates this
   energy at much lower frequencies, producing a short optical flash
   of 15th magnitude or brighter.

   \end{abstract}

%

\section{Introduction}

In the internal-external scenario, the GRB is produced by internal
shocks while the afterglow is produced by the interaction of the flow
with the ISM.  The original fireball model was invoked to explain the
Gamma-Ray Bursts (GRBs) phenomena. It requires extreme relativistic
motion, with a Lorentz factor $\gamma \gsim 100$.

The afterglow observations, which fit the theory rather well, are
considered as a confirmation of the fireball model.  However, the
current afterglow observations, which detect radiation from several
hours after the burst onwards, do not probe the initial extreme
relativistic conditions. By the time of the present observations,
several hours after the burst, the Lorentz factor is less than $\sim
10$, and it is independent of the initial Lorentz factor.

Afterglow observations, a few seconds after
the burst, can provide the missing information concerning the
initial phase and prove the internal shock scenario. 
Such rapid observations are possible, in principle, with future
missions (Kulkarni and Harrison, Private communication).

\section{The Forward Shock}
   
The synchrotron spectrum from relativistic electrons that are
continuously accelerated into a power law energy distribution is
always given by four power law segments, separated by three critical
frequencies: $\nu _{sa}$ the self absorption frequency, $\nu _{c}$ the
cooling frequency and $\nu _{m}$ the characteristic synchrotron
frequency. 

Using the relativistic shock jump conditions and assuming that the
electrons and the magnetic field acquire fractions $\epsilon _{e}$
and $\epsilon _{B}$ of the equipartition energy, we obtain:
\begin{equation}
\label{num}
\nu _{m}=1.1\times 10^{19}{\rm Hz}\left( \frac{\epsilon _{e}}{0.1}\right)
^{2}\left( \frac{\epsilon _{B}}{0.1}\right) ^{1/2}(\frac{\gamma}{300}%
)^{4}n_{1}^{1/2}.
\end{equation}
\begin{equation}
\nu _{c}=1.1\times 10^{17}{\rm Hz}\left( \frac{\epsilon _{B}}{0.1}\right)
^{-3/2}\left( \frac{\gamma }{300}\right) ^{-4}n_{1}^{-3/2}t_{s}^{-2},
\end{equation}
\begin{equation}
F_{\nu ,\max }=220{\rm \mu J}D_{28}^{-2}\left( \frac{\epsilon _{B}}{0.1}%
\right) ^{1/2}\left( \frac{\gamma }{300}\right) ^{8}n_{1}^{3/2}t_{s}^{3}
\end{equation}
\begin{equation}
\label{nusa}
\nu _{sa}=220{\rm GHz}\left( \frac{\epsilon _{B}}{0.1}\right) ^{6/5}\left(
\frac{\gamma }{300}\right) ^{28/5}n_{1}^{9/5}t_{s}^{8/5}
\end{equation}
These scalings generalize the adiabatic scalings obtained by
Sari, Piran \& Narayan (1998) to an arbitrary hydrodynamic evolution of
$\gamma(t)$.

For typical parameters, $\nu_c < \nu_m$, so fast cooling occurs. 
The spectrum of fast cooling electrons is described by four power
laws: (i) For $\nu <\nu _{sa}$ self absorption is important and
$F_{\nu }\propto \nu ^{2}$. (ii) For $\nu _{sa}<\nu <\nu _{c}$ we have
the synchrotron low energy tail $F_{\nu }\propto \nu ^{{1/3}}$. (iii)
For $\nu _{c}<\nu <\nu _{m}$ we have the electron cooling slope
$F_{\nu }\propto \nu ^{-1/2}$. (iv) For $\nu >\nu _{m}$ 
$F_{\nu }\propto \nu ^{-p/2}$, where $p$ is the index of 
the electron power law distribution.

In the early afterglow, the Lorentz factor is initially
constant. After this phase, the evolution can be of two types (Sari 1997). 
Thick shells,
which correspond to long bursts, begin to decelerate with
$\gamma(t)\sim t^{-1/4}$. Only later there is a transition to
deceleration with $\gamma(t)\sim t^{-3/8}$. The light
curves for such bursts can be obtained by substituting these scalings
in equations
\ref{num}-\ref{nusa}. However, for these long bursts, 
the complex internal shocks GRB signal would overlap, 
the smooth external shock afterglow signal. 
The separation of the observations to GRB and early
afterglow would be rather difficult.

For thin shells, that correspond to short bursts, there is no
intermediate stage of $\gamma(t)\sim t^{-1/4}$. 
There is a single transition, at the time $t_\gamma=(
3E/32\pi \gamma _{0}^{8}nm_{p}c^{5}) ^{1/3}$, 
from a constant velocity to self-similar deceleration with
$\gamma(t)\sim t^{-3/8}$. The possible light curve
are illustrated in figure 1. As the intial afterglow peaks
several dozen seconds after the GRB there should be no difficulty to detect it.

The detection of delayed emission which fits the light curves of
figure 1, would enable us to determine $t_\gamma$. Using $t_\gamma$ we
could proceed to estimate the initial Lorentz factor:
\begin{equation}
\gamma _{0}= 
240E_{52}^{1/8}n_{1}^{1/8}\left( t_\gamma/10{\rm s}\right) ^{-3/8}.
\label{gammaoft}
\end{equation}
If the second peak of GRB 970228, delayed by 35s, is indeed the afterglow 
rise, then $\gamma_0 \sim 150$ for this burst.

%
   \begin{figure}
     \epsfxsize=8.8cm \epsfbox{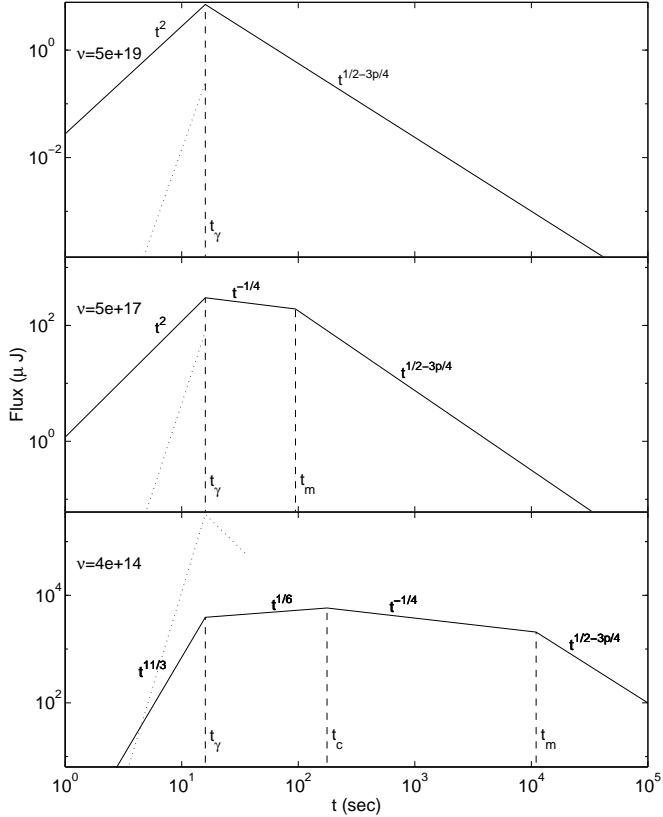}
      \caption[]{Light curves of the forward (solid) and reverse (dashed)
                shocks in three energy bands.
              }
         \label{FigVibStab}
   \end{figure}
%

\section{The Reverse Shock and the Optical Flash}
There are many attempts to detect
early optical emission and there is a good chance that this emission
will be observed in the near future.  A strong 5th magnitude optical
flash would have been produced if the fluence of a moderately strong
GRBs, $10^{-5}$erg/s/cm$^2$ would have been released on a time scale
of 10s in the optical band.  Even a small fraction of this will be
easily observed.  It is important, therefore, to explore the expected
optical emission from the GRB and the early afterglow.

During the GRB and the initial emission from the forward shock the
emission peaks in $\gamma$-rays, and only an extremely small fraction is
emitted in the optical band. For example, the prompt optical flash from the
GRB would be of 21st magnitude if the flux drops according to the
synchrotron low energy tail of $F_\nu \sim \nu^{1/3}$.

A considerably stronger flux is obtained from the reverse shock. The
reverse shock contains, at the time it crosses the shell, a comparable
amount of energy to the forward shock. However, its effective
temperature is significantly lower (typically by a factor of $\gamma
\sim 300$) than that of the forward shock.  The resulting peak
frequency is therefore lower by $\gamma^2 \sim 10^5$.  A more detailed
calculation shows that the reverse shock frequency is
\begin{equation}
\nu _{m}=1.2\times 10^{14}\left( \frac{\epsilon _{e}}{0.1}\right)
^{2}\left( \frac{\epsilon _{B}}{0.1}\right) ^{1/2}(\frac{\gamma _{0}}{300}%
)^{2}n_{1}^{1/2} .
\end{equation}
The cooling frequency is similar to that of the forward shock, 
since both have the same magnetic field and the same Lorentz factor. 
Using the parameters obtained by Granot, Piran and Sari (1998), 
from the afterglow
of GRB 970508, and using $\gamma_0=300$ we get for the reverse shock
$\nu_c=3\times 10^{16}$Hz and $\nu_m=3\times 10^{14}$Hz leading to an 8th 
magnitude flash. With higher initial Lorentz factor of $\gamma_0=10^4$ the flash
drops to 13th magnitude. 
Inverse Compton cooling, if exists, can reduce the flux by $\sim 2$ magnitudes,
while self absoption can influence only very short bursts with small surface
area. Therefor, quite concervatively, the optical flash
should be stronger than 15th magnitude and should be soon
seen with modern experiments. The reverse shock signal is very short living.
After the shock crosses the shell, no new electrons are injected 
and there is no emission above $\nu_c$. Moreover, $\nu_c$ drops fast 
with time as the shell's material cools adiabatically.

\section{Discussion}

The early afterglow multi-wavelength radiation could provide
interesting and invaluable information on the extreme relativistic
conditions that occur at this stage. The initial emission from the forward
shock, which continues later as the observed afterglow signals is in
the $\gamma$-rays or x-rays. The reverse shock emission
could have a strong, short living,
optical component which we expect to be brighter than 15th
magnitude. Some of the current basic ideas concerning the fireball
models should be revised if such signals will not be seen by new
detectors that should become operational in the near future.

\begin{acknowledgements}
This research was supported by the US-Israel BSF 95-328 and
by a grant from the Israeli Space Agency. R.S. thanks the
Sherman Fairchild Foundation for support.
\end{acknowledgements}

\end{document}